\documentclass[prd,preprint,showpacs,superscriptaddress]{revtex4-1}
\usepackage{amsmath}
\usepackage{latexsym}
\usepackage{amsfonts}
\usepackage{graphicx}
\usepackage{hyperref}

\begin{document}

\title{The general property of dynamical quintessence field}
\author{Yungui Gong}
\email{yggong@mail.hust.edu.cn}
\affiliation{MOE Key Laboratory of Fundamental Quantities Measurement, School of Physics, Huazhong University of Science and Technology, Wuhan 430074, China}
\affiliation{Institute of Theoretical Physics, Chinese Academy of Sciences, Beijing 100190, China}

\begin{abstract}
We discuss the general dynamical behaviors of quintessence field, in particular,
the general conditions for tracking and thawing solutions are discussed. We explain what the tracking solutions mean
and in what sense the results depend on the initial conditions. Based on the definition of tracking solution,
we give a simple explanation on the existence of
a general relation between $w_\phi$ and $\Omega_\phi$ which is independent of the initial conditions for the tracking solution.
A more general tracker theorem which requires large initial values of the roll parameter is then proposed.
To get thawing solutions, the initial value
of the roll parameter needs to be small. The power-law and pseudo-Nambu Goldstone
boson potentials are used to discuss the tracking
and thawing solutions. A more general $w_\phi-\Omega_\phi$ relation is derived for the thawing solutions. Based
on the asymptotical behavior of the $w_\phi-\Omega_\phi$ relation,
the flow parameter is used to give an upper limit
on $w_\phi'$ for the thawing solutions. If we use the observational constraint $w_{\phi 0}<-0.8$ and $0.2<\Omega_{m0}<0.4$,
then we require $n\lesssim 1$ for the inverse power-law potential $V(\phi)=V_0(\phi/m_{pl})^{-n}$ with tracking solutions and
the initial value of the roll parameter $|\lambda_i|<1.3$ for the potentials with the thawing solutions.
\end{abstract}

\maketitle

\section{Introduction}

The recent cosmic acceleration observed by type Ia supernova data \cite{acc0,*hzsst98,*scpsn98}
was usually explained by introducing a dynamical scalar field called quintessence \cite{peebles88,*wetterich88,quintessence,track1}.
More general dynamical scalar field models such as phantom \cite{phantom}, quintom \cite{Feng:2004ad,*Feng:2004ff,*Guo:2004fq},
tachyon \cite{Sen:2002in,*Sen:2002nu,*Padmanabhan:2002cp} and k-essence \cite{ArmendarizPicon:2000dh}
were also proposed. For a recent review of dark energy,
please see Ref. \cite{sahni00,*copelandde,*Nojiri:2006ri,*Padmanabhan:2007xy,*limiaode,*Nojiri:2010wj,*Bamba:2012cp}.

For a dynamical scalar field $\phi$ with the potential $V(\phi)$ in
the flat Friedmann-Lema\^{i}tre-Robertson-Walker universe with the metric
$ds^2=-dt^2+a^2(t)(dr^2+r^2d\theta^2+r^2\sin^2\theta d\phi^2)$, its energy density and pressure
are $\rho_\phi=\dot\phi^2/2+V(\phi)$ and $p_\phi=\dot\phi^2/2-V(\phi)$, where $\dot\phi=d\phi/dt$.
The scalar field rolls down a very shallow potential
while its equation of state $w_\phi=p_\phi/\rho_\phi$ approaches $-1$ and it starts to dominate
the Universe recently. Because the scalar field catches up the background only
recently and the current value of its equation of state parameter is around $-1$,
$w_\phi$ does not change too much in the redshift range $0\le z<1$ for most scalar fields,
so the time variation of $w_\phi$ is bounded for the thawing and freezing models \cite{Caldwell:2005tm,Barger:2005sb,Scherrer:2005je,Linder:2006sv,chiba06,*chiba06a,Ali:2009mr,Cahn:2008gk,Lopez:2011ur,Chen:2013vba}.
In general, the evolution of scalar field depends on the initial conditions. However,
the attractor solutions and the tracking solutions are independent of the initial conditions
\cite{Ferreira:1997au,*Ferreira:1997hj,ejcetal98,Liddle:1998xm,
track2,Brax:1999yv,UrenaLopez:2000aj,Bludman:2001iz,Dodelson:2001fq,
Johri:2000yx,*Johri:2001wm,Rubano:2003et,Watson:2003kk,Aguir:2004xd,
Sahlen:2006dn,Huterer:2006mv,Fang:2008fw,Szydlowski:2013sma,delCampo:2013vka,Roy:2013wqa}.
In particular, the tracker field $\phi$ tracks below the background density for most of the history of the Universe
until it starts to dominate recently for a wide range of initial conditions, and there exists
a relation between $w_\phi$ and the fractional energy density $\Omega_\phi=8\pi G\rho_\phi/(3H^2)$ today,
where the Hubble parameter $H(t)=\dot a/a$.
There also exists a general $w_\phi-\Omega_\phi$ relation for the thawing solutions which
is well approximated by some analytical expressions \cite{Robert2008,Robert,Sourish,Crittenden:2007yy,Dutta:2008qn,
Chiba:2009nh,Gupta:2009kk,Chiba:2009sj,Sen:2009yh,delCampo:2010fg,Gong:2013bn}.
In this Letter, we will discuss the general dynamics such as the $w_\phi-\Omega_\phi$ relation
and the bound on $w'_\phi=dw_\phi/d\ln a$ of the tracking and thawing fields.
We use the power-law potential and the pseudo-Nambu Goldstone boson (PNGB)
potential \cite{Hill:1987bm,Hill:1988bu,Freese:1990rb,Frieman:1995pm,Coble:1996te} as examples
to illustrate the general dynamical behaviors of tracking and thawing fields.

If the Universe is filled with the quintessence field and the background matter with the
equation of state $w_b=[(1/3)a_{eq}/a]/[1+a_{eq}/a]$, where $a_{eq}=1/3403$ \cite{Ade:2013zuv} is the scale factor $a(t)$
at the matter-radiation equality, then
in terms of the dimensionless variables,
\begin{equation}
\label{vareq1}
x=\frac{\phi'}{\sqrt{6}}=\frac{1}{\sqrt{6}}\frac{d\phi}{d\ln a},\quad y=\sqrt{\frac{V}{3H^2}},
\quad \lambda=-\frac{V_{,\phi}}{V}=-\frac{1}{V}\frac{dV}{d\phi},\quad
\Gamma=\frac{V V_{,\phi\phi}}{V_{,\phi}^2},
\end{equation}
the cosmological equations are
\begin{gather}
\label{dynaeq1}
x'=\sqrt{\frac{3}{2}}\lambda  y^2+\frac{3}{2}x(x^2-y^2-1)+\frac{3}{2}w_b x(1- x^2-y^2),\\
\label{dynaeq2}
y'=-\sqrt{\frac{3}{2}}\lambda xy+\frac{3}{2}y(1+x^2-y^2)+\frac{3}{2}w_b y(1-x^2-y^2),\\
\label{dynaeq3}
\lambda'=-\sqrt{6}\lambda^2(\Gamma-1)x.
\end{gather}
The fractional energy density and the equation of state of the scalar field are
\begin{equation}
\label{phyeq1}
\Omega_\phi=x^2+y^2,\quad w_\phi=\frac{x^2-y^2}{x^2+y^2}.
\end{equation}
Using the fractional energy density $\Omega_\phi$ and the parameter $\gamma=1+w$,
Eqs. (\ref{dynaeq1})-(\ref{dynaeq3}) become
\begin{gather}
\label{dynaeq1a}
\Omega'_\phi=3(\gamma_b-\gamma_\phi)\Omega_\phi(1-\Omega_\phi),\\
\label{dynaeq2a}
\gamma'_\phi=(2-\gamma_\phi)(-3\gamma_\phi+|\lambda|\sqrt{3\gamma_\phi\Omega_\phi}),\\
\label{dynaeq3a}
\lambda'=-\sqrt{3\gamma_\phi\Omega_\phi}\lambda |\lambda| (\Gamma-1).
\end{gather}

From Eq. (\ref{dynaeq2a}), we get a lower limit $w'_\phi\ge -3(1+w_\phi)(1-w_\phi)$.
If the tracker parameter $\Gamma$ can be expressed as a function of the roll parameter $\lambda$, then the above system (\ref{dynaeq1a})-(\ref{dynaeq3a})
becomes an autonomous system.
In this case, we have additional critical point $\Omega_{\phi c}=1$ and $\gamma_c=0$ which
is absent in the system (\ref{dynaeq1})-(\ref{dynaeq3}), where the subscript $c$ means the critical point.
At the point $x=0$ and $y=1$, the
transformation from $(x,\ y)$ to $(\Omega_\phi,\ \gamma)$ is singular, so we get different critical points.
Only when $\lambda_c=0$, the point $x=0$ and $y=1$ is the critical point of the system (\ref{dynaeq1})-(\ref{dynaeq3}).
For the exponential potential, the point $x=0$ and $y=1$  is not a critical point for the system (\ref{dynaeq1})-(\ref{dynaeq3})
and the critical point $\Omega_{\phi c}=1$ and $\gamma_c=0$  for the system (\ref{dynaeq1a})-(\ref{dynaeq3a}) is not
a stable point.

If we use the flow parameter $F=\gamma_\phi/(\Omega_\phi \lambda^2)$ \cite{Cahn:2008gk}, then Eq. (\ref{dynaeq2a})
can be written as
\begin{equation}
\label{wvareq1}
\gamma_\phi'=3\gamma_\phi(2-\gamma_\phi)(-1+1/\sqrt{3F}).
\end{equation}
To understand the general dynamics of the quintessence, it is useful to
use the function $\beta=\ddot\phi/(3H\dot\phi)$ \cite{Cahn:2008gk,Chiba:2009sj},
\begin{equation}
\label{betaeq1}
\beta=-1+\frac{1-w_\phi}{\sqrt{12F}}=\frac{1}{2}[\Omega_\phi\gamma_\phi
+(1-\Omega_\phi)\gamma_b]-\frac{\beta'}{3(1+\beta)}-\frac{V_{,\phi\phi}}{9(1+\beta)H^2}.
\end{equation}
For the thawing solution, the quintessence field rolls down the potential very slowly,
$V_{,\phi\phi}\approx 0$ and $\beta$ is almost a constant, so $\beta\approx \gamma_b/2$
at early time when $\Omega_\phi\approx 0$ and $w_\phi\approx -1$ \cite{Chiba:2009sj}.

\section{Tracker solution}

The energy density of the tracker field $\phi$ tracks below the background density for most of
the history of the Universe, it starts to dominate the energy density only recently and then drives
the cosmic acceleration. The tracker fields have attractor-like solutions in the sense
that they rapidly converge to a common cosmic evolutionary track of $\rho_\phi(t)$
and $w_\phi(t)$ for a very wide range of initial conditions, so the tracking solutions
are extremely insensitive to the initial conditions \cite{track1,track2}. Furthermore,
an important relation between $w_\phi$ and $\Omega_\phi$ today was found for the tracker field.
When the tracker field enters the tracking solution, it satisfies the tracker condition \cite{track2}
\begin{equation}
\label{trkcond}
\gamma_\phi=1+w_\phi=\frac{1}{3}\lambda^2\Omega_\phi,
\end{equation}
thus this condition is the initial condition of tracking solution.
In other words,
the initial condition for the tracking solution reads $F=1/3$.
From Eq. (\ref{dynaeq2a}), we see that $\gamma_\phi'=0$ when the tracker
condition is satisfied, so it is possible that $w_\phi$ stops varying.
On the other hand, the quintessence field satisfies the tracker equation \cite{track2,Rubano:2003et}
\begin{equation}
\label{trackereq}
\Gamma-1=\frac{3(w_b-w_\phi)(1-\Omega_\phi)}{(1+w_\phi)(6+\tilde x')}
-\frac{(1-w_\phi)\tilde x'}{2(1+w_\phi)(6+\tilde x')}-\frac{2\tilde x''}{(1+w_\phi)(6+\tilde x')^2},
\end{equation}
where  $\tilde x=\ln[(1+w_\phi)/(1-w_\phi)]$ and $\tilde x'=d\ln \tilde x/d\ln a$.
For the tracking solution, $w_\phi$ is nearly constant, so $\tilde x'\approx \tilde x''\approx 0$,
and we get
\begin{equation}
\label{trksol1}
w_\phi\approx\frac{w_b(1-\Omega_\phi)-2(\Gamma-1)}{2\Gamma-1-\Omega_\phi}< w_b,\quad (\Gamma>1).
\end{equation}
If $\Omega_\phi\approx 0$ when the tracker condition (\ref{trkcond}) is satisfied,
then
\begin{equation}
\label{trksol2}
w_\phi=w_\phi^{trk}=\frac{w_b-2(\Gamma-1)}{2\Gamma-1},
\end{equation}
and $\beta=-\gamma_b/2(2\Gamma-1)$ are approximately constants if the tracker parameter
$\Gamma$ is nearly constant,
$\Omega_\phi\propto a^{6\gamma_b(\Gamma-1)/(2\Gamma-1)}$ increases with time and
$\lambda^2\approx 3(1+w_\phi^{trk})/\Omega_\phi$ decreases with time.
For the tracker field, $V_{,\phi\phi}/H^2$ is not negligible, so $\beta\neq \gamma_b/2$. In fact,
$V_{,\phi\phi}/H^2$ is a constant for the exponential potential when the attractor is reached.

If $\Omega_\phi$ is not small or the tracker parameter changes rapidly
when the tracker condition (\ref{trkcond}) is satisfied,
then $w_\phi$ won't keep to be a time independent constant and it
decreases with time while $\Omega_\phi$ increases to 1,
so the scalar field does not track the background and
Eq. (\ref{trksol2}) does not hold when the tracker condition (\ref{trkcond}) is satisfied,
but the scalar field has the the freezing behavior with $w_\phi\rightarrow -1$ asymptotically.
Therefore, both the tracker condition (\ref{trkcond}) and Eq. (\ref{trksol2})
will be violated when $\Omega_\phi$ is not negligible or $\Gamma$ changes rapidly,
and $w_\phi$ keeps decreasing.

For the tracker field, the tracking solution at late times has the property
that $\gamma_\phi\rightarrow 0$ and $\Omega_\phi\rightarrow 1$,
so $\gamma_\phi$ should decrease with time while $\Omega_\phi$ increases with time.
When $\gamma_\phi$ reaches the background value $\gamma_b$, and $\lambda^2$ decreases
to the value $\lambda^2=3\gamma_\phi/\Omega_\phi$,
then we reach the tracker condition. After that, $\gamma_\phi$ decreases toward to zero and $\Omega_\phi$ increases toward 1.
From Eq. (\ref{dynaeq2a}), we know that we should keep $|\lambda|<\sqrt{3\gamma_\phi/\Omega_\phi}$
in order that $\gamma'_\phi<0$, therefore $|\lambda|$ should
decrease with time and $\lambda\rightarrow 0$ when $\gamma_\phi\rightarrow 0$. For any quintessence field
rolling down its potential, $|\lambda|$ does not increase with time is
equivalent to $\Gamma\ge 1$ as easily seen from Eq. (\ref{dynaeq3a}).

\begin{figure}[htp]
\centerline{\includegraphics[width=0.6\textwidth]{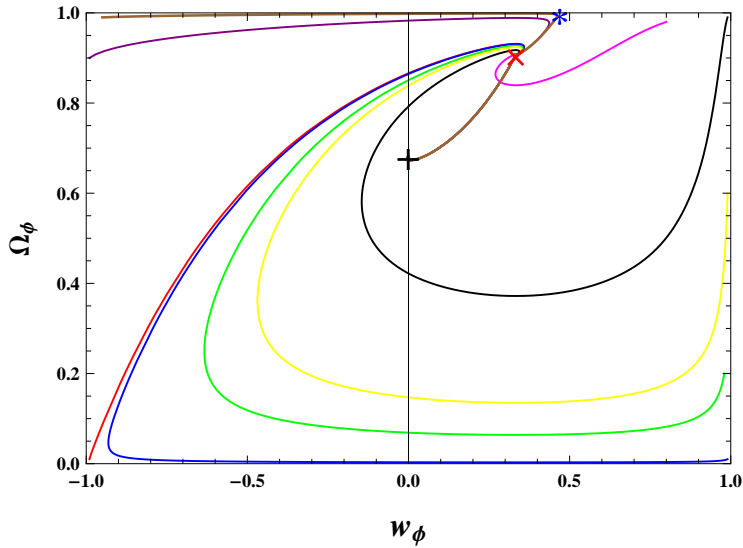}}
\caption{The phase diagram of $w_\phi$ and $\Omega_\phi$ for the exponential potential with $\lambda=2.1$.
"*" corresponds to the saddle node $\Omega_\phi=1$ and $\gamma_\phi=\lambda^2/3$, "$\times$" and "+" correspond to the stable points
$\gamma_\phi=\gamma_b$ and $\Omega_\phi=3\gamma_\phi/\lambda^2$ with $\gamma_b=4/3$ and $\gamma_b=1$ respectively.}
\label{exptrak2}
\end{figure}

For the exponential potential, $\lambda$ is a constant
and $\Gamma=1$. If $\lambda$ is small, then eventually $\gamma_\phi$ will decrease to be less than $\gamma_b$, and
$\Omega_\phi$ will quickly increase to be 1. In particular, if $\lambda^2<3\gamma_b$,
then the system will reach the attractor solution with
$\Omega_\phi=1$ and $\gamma_\phi=\lambda^2/3$. If $\lambda$ is big, i.e., $\lambda^2\ge 3\gamma_b$,
then the attractor solution is $\gamma_\phi=\gamma_b=\lambda^2\Omega_\phi/3$. Since
the above attractors satisfy the tracker condition (\ref{trkcond}), so both of
them are also tracking solutions. In Fig. \ref{exptrak2}, we show the phase
diagram for the exponential potential with $\lambda=2.1$.
The original tracking solution found in \cite{track2} is independent of the value of $\lambda$
which is in contradiction with the results for the exponential potential. The contradiction was
then resolved in \cite{Rubano:2003et} by deriving the correct tracker equation (\ref{trackereq}).

With the dynamical Eqs. (\ref{dynaeq1a})-(\ref{dynaeq3a}), we
can understand the general dynamical evolution of the tracker field as follows:
(a) Initially if $\Omega_{\phi i}$ is not too small or $\lambda_i$ is large enough so that $\lambda_i^2>3\gamma_{\phi i}/\Omega_{\phi i}$,
where the subscript $i$ means the initial value,
then $\gamma_\phi$ will increase toward 2 independent of the initial value of $w_{\phi}$.
Once $\gamma_\phi>\gamma_b$, $\Omega_\phi$ will decrease. When $\Omega_\phi$ decreases to be small enough
so that $\lambda^2<3\gamma_\phi/\Omega_\phi$, $\gamma_\phi$ decreases toward $-1$ and $\Omega_\phi$
starts to increase once $\gamma_\phi<\gamma_b$.
Even if $\lambda$ decreases, it will overtake $3\gamma_\phi/\Omega_\phi$ when $\gamma_\phi\rightarrow 0$, then $\gamma_\phi$
increases again. Once $\gamma_\phi$ increases away from zero, we will have $\lambda^2<3\gamma_\phi/\Omega_\phi$
and the tracker behavior of $w_\phi$
will be reached, maybe after several oscillations. If $\lambda_i$ is small, then
to satisfy the tracker condition, $\Omega_{\phi i}$ cannot be too small
and $\Omega_\phi$ may reach $1$ very quickly and the tracker solution with nearly constant $w_\phi$ will not appear.
(b) Initially if $\Omega_{\phi i}$ or $\lambda_i$ is small so that $\lambda_i^2<3\gamma_{\phi i}/\Omega_{\phi i}$,
then $\gamma_\phi$ decreases toward 0 and $\Omega_\phi$ starts to increase once $\gamma_\phi<\gamma_b$
independent of the initial value $\gamma_{\phi i}$. After that, the dynamics is similar to that in case (a).
In this case, even though $\Omega_{\phi i}$ is small, but it always increases if $\lambda_i$ is small, $\Omega_\phi$
will reach 1 soon and the tracker solution with nearly constant $w_\phi$ does not appear.
Of course, the current values of $\Omega_\phi$, $\lambda$ and $w_\phi$ depend on their initial values.
From these
analyses, we conclude that if the initial value of $\lambda$ is small, then no tracker solution with nearly constant $w_\phi$ appears
because $\Omega_\phi$ reaches 1 too soon. For small $\lambda_i$,
once $\Omega_\phi$ reaches 1, the equation for $\Omega_\phi$ decouples from the dynamical system (\ref{dynaeq1a})-(\ref{dynaeq3a}).
Therefore, the solution to Eqs. (\ref{dynaeq2a}) and (\ref{dynaeq3a}) gives a relation between $w_\phi$ and $\lambda$.
Since $w_\phi$ approaches $-1$ asymptotically, the relation is universal if $\Gamma$ is a function of $\lambda$
in the sense that it does not depend on the initial conditions, therefore the $w_\phi-w_\phi'$ relation is also
universal when $w_\phi\rightarrow -1$.

For the tracking solution, the conditions (\ref{trkcond}) and
(\ref{trksol2}) are the initial conditions, so for the same initial value of $\Omega_\phi$ at the start
of the tracking solution, the trajectories
of $w_\phi$, $\Omega_\phi$ an $\lambda$ will be the same, that is why we have the same
$w_\phi-\Omega_\phi$ trajectory for the tracking solution independent of the initial conditions.
However, the exact values of $w_\phi$ and $\Omega_\phi$ at a moment (for instance, at the present) still depend on
the initial conditions. {\em We refer the tracking solutions
as those solutions which have a common $w_\phi-\Omega_\phi$ trajectory for a wide range of initial conditions,
technically, the tracking solutions satisfy the two conditions (\ref{trkcond}) and
(\ref{trksol2}) initially}.

In \cite{Lopez:2011ur}, the author rephrased the tracker theorem as:
the tracker property appears for any scalar field model in which
the roll parameter $\lambda$ is capable of taking on large initial values in the early Universe. If $\lambda$ decreases with time,
then $\lambda$ is capable of taking on large initial values in the early Universe. However,
for the power-law potential $\phi^\alpha$ with $\alpha>0$, the roll parameter
$\lambda$ can be large if we start from small $\phi$, and there is no tracking solution. Furthermore,
it is not clear how large the initial value should be.
Therefore, we propose the tracker
theorem as:  {\em the tracker behavior appears for any quintessence field in which
the roll parameter $|\lambda|$ does not increase with time and the initial value
of $\lambda$ should be big enough so that $\Omega_\phi$ is still negligible when
the tracker condition (\ref{trkcond}) is satisfied}. The new definition of the tracking
solution and the tracker theorem proposed here are parts of the main results of this Letter.

\section{Power-law potential}

In this section, we use the power-law potential as an example to explicitly show
the analyses presented in the previous section. Here we focus on the tracking and thawing
behaviors and the bound on $w_\phi'$.
For the power-law potential $V(\phi)=V_0(\phi/m_{pl})^\alpha$ with the energy
scale $V_0\sim (10^{-3}\ {\rm eV})^4$ and the Planck mass $m_{pl}=1/(8\pi G)^{1/2}$,
the tracker parameter $\Gamma=1-\alpha^{-1}$,
so the dynamical system becomes an autonomous system. The dynamical analysis of the
system (\ref{dynaeq1})-(\ref{dynaeq3}) with $f(\lambda)=-1/\alpha$ was carried out in \cite{Fang:2008fw}.
The dynamical system (\ref{dynaeq1})-(\ref{dynaeq3}) has the following critical points: $(x_c, y_c, \lambda_c)=(\pm 1, 0,0)$,
$(x_c, y_c)=(0,0)$ with $\lambda$ arbitrary, and $(x_c, y_c, \lambda_c)=(0, 1,0)$. Only the critical point $(x_c, y_c, \lambda_c)=(0, 1,0)$
can be stable if $f(\lambda=0)>0$. For the inverse power-law potential,
$f(0)=-1/\alpha>0$, so the critical point $(x_c, y_c, \lambda_c)=(0, 1,0)$
which corresponds to the solution $\Omega_{\phi c}=1$ and $\gamma_{\phi c}=0$ is a stable point.
From this analysis, we know that $\lambda$
will decrease to zero for the tracking solution.

\subsection{Tracking solution}

If we use the dynamical system (\ref{dynaeq1a})-(\ref{dynaeq3a}), the critical points are: $\Omega_{\phi c}=0$, $\gamma_{\phi c}=0$
or $\gamma_{\phi c}=2$ with arbitrary $\lambda$; $\Omega_{\phi c}=1$ and $\gamma_{\phi c}=0$ with arbitrary $\lambda$; and
$(\Omega_{\phi c},\gamma_{\phi c},\lambda_c)=(1,2,0)$. The critical point $(\Omega_{\phi c},\gamma_{\phi c},\lambda_c)=(1,0,0)$ is a stable point.
For the critical point $(\Omega_{\phi c},\gamma_{\phi c},\lambda_c)=(1,0,0)$,
the linear approximation of the system (\ref{dynaeq1a})-(\ref{dynaeq3a}) is
\begin{gather}
\label{pwrlinear1}
\delta \Omega_\phi'=-3\gamma_b\delta\Omega_\phi,\\
\label{pwrlinear2}
\delta\gamma_\phi'=-6\delta\gamma_\phi,\\
\label{pwrlinear3}
\delta\lambda'=0.
\end{gather}
One of the eigenvalues is 0. To analyze the stability of the system, we need to understand the stability of equation (\ref{dynaeq3a})
for the critical point $(\Omega_{\phi c},\gamma_{\phi c},\lambda_c)=(1,0,0)$
by using the center manifold theorem \cite{hkkhalil02}. So we need to solve
the following equation \cite{hkkhalil02}
\begin{equation}
\label{zeropt1}
\frac{d\gamma_\phi}{d\lambda}\sqrt{3\gamma_\phi\Omega_\phi}\lambda^2/\alpha=(2-\gamma_\phi)(-3\gamma_\phi+\lambda\sqrt{3\Omega_\phi\gamma_\phi}).
\end{equation}
Let $\gamma_\phi(\lambda)=\gamma_2\lambda^2$, up to the order of $\lambda^2$, we get $\gamma_2=1/3$. Substituting
$\Omega_\phi=1$ and $\gamma_\phi=\lambda^2/3$  into Eq. (\ref{dynaeq3a}), we get
\begin{equation}
\label{zeropt2}
\lambda'=\lambda^3/\alpha.
\end{equation}
The system is stable if $\alpha<0$, so the critical point $(\Omega_{\phi c},\gamma_c,\lambda_c)=(1,0,0)$
of the dynamical system (\ref{dynaeq1a})-(\ref{dynaeq3a}) is a stable point for the inverse power-law potential.
From the above analysis, we see that asymptotically $\gamma_\phi=\lambda^2\Omega_\phi/3$ to the leading order,
so this stable critical point corresponds to the late time tracking solution.
In other words, the flow parameter
starts with the value $F=1/3$ and approaches the value $F=1/3$ asymptotically for the tracking solution.
The function $\beta$ starts with $\beta=\alpha\gamma_b/2(2-\alpha)$ and increases to $\beta=0$ asymptotically.

\begin{figure}[htp]
\begin{tabular}{cc}
\includegraphics[width=0.4\textwidth]{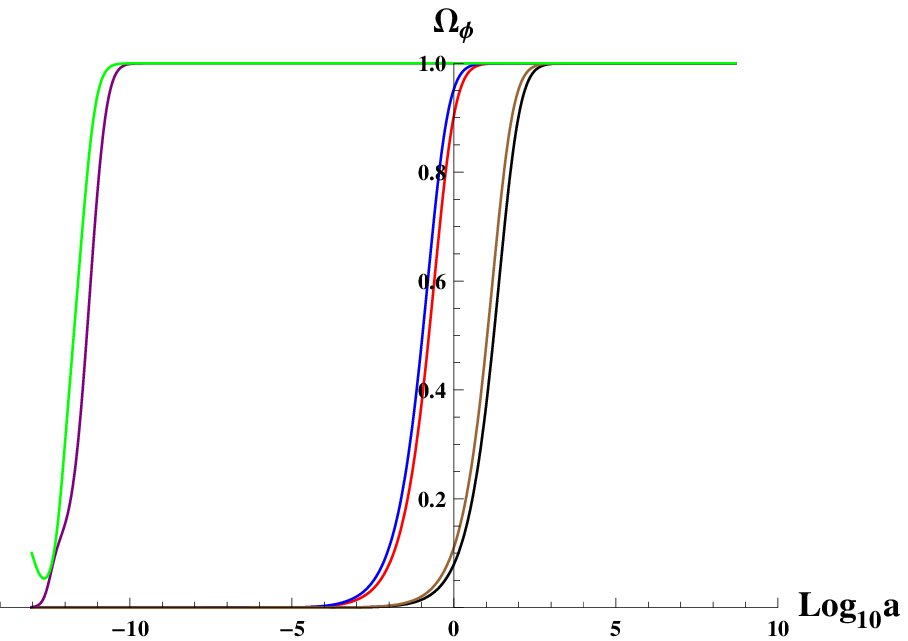}&\includegraphics[width=0.4\textwidth]{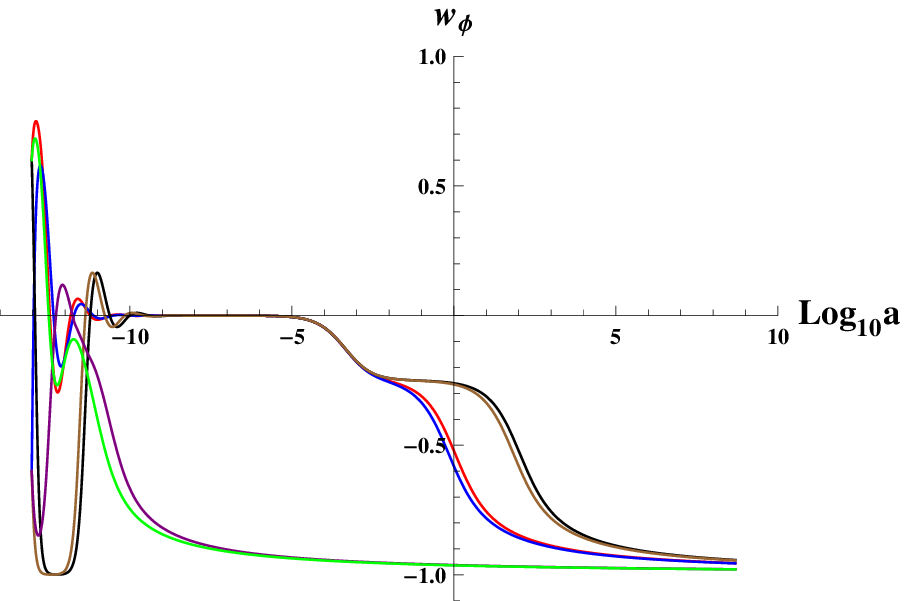}
\end{tabular}
\caption{The time evolutions of $\Omega_\phi$ and $w_\phi$ for the inverse power-law potential $V(\phi)\sim 1/\phi^6$.
The arbitrary initial time
$\ln a_i=-30$ was chosen for computational convenience.
The initial condition for the red line is $\Omega_{\phi i}=10^{-11}$, $w_{\phi i}=0.6$ and $\lambda_i=1.2\times 10^6$.
The initial condition for the blue line is $\Omega_{\phi i}=10^{-11}$, $w_{\phi i}=-0.6$ and $\lambda_i=1.2\times 10^6$.
The initial condition for the black line is $\Omega_{\phi i}=10^{-17}$, $w_{\phi i}=0.6$ and $\lambda_i=1.2\times 10^6$.
The initial condition for the brown line is $\Omega_{\phi i}=10^{-17}$, $w_{\phi i}=-0.6$ and $\lambda_i=1.2\times 10^6$.
The initial condition for the purple line is $\Omega_{\phi i}=10^{-3}$, $w_{\phi i}=-0.6$ and $\lambda_i=10$.
The initial condition for the green line is $\Omega_{\phi i}=0.1$, $w_{\phi i}=0.6$ and $\lambda_i=10$.}
\label{pwrtrak1}
\end{figure}

\begin{figure}[htp]
\begin{tabular}{cc}
\includegraphics[width=0.4\textwidth]{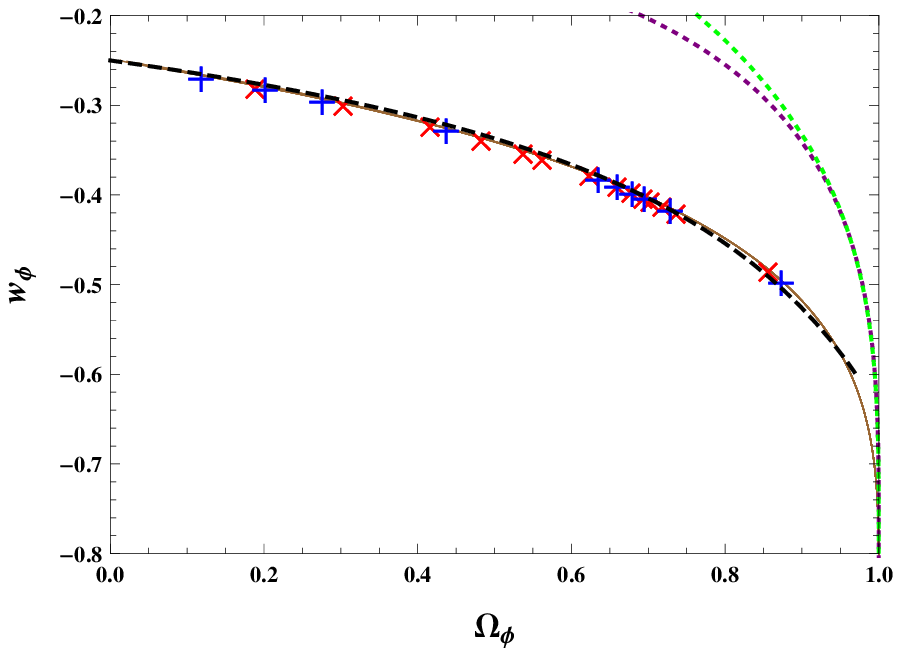}&\includegraphics[width=0.4\textwidth]{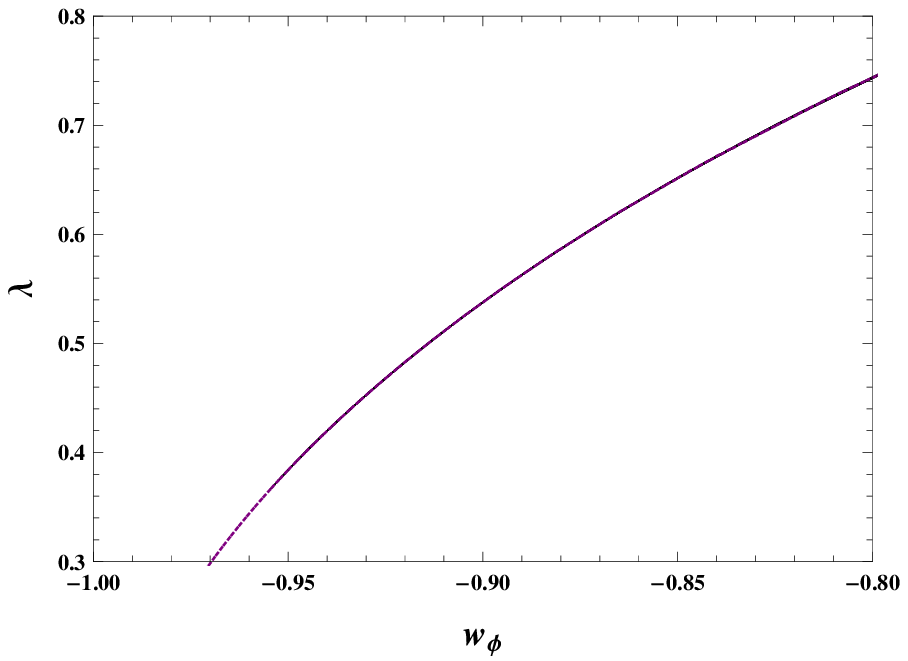}
\end{tabular}
\caption{The left panel shows the $w_\phi-\Omega_\phi$ trajectories of the evolutions of $w_\phi$ and $\Omega_\phi$
shown in Fig. \ref{pwrtrak1},
and the right panel shows $\lambda$ versus $w_\phi$ for the inverse power-law
potential $V(\phi)\sim 1/\phi^6$ with the tracking behavior.
For $\lambda_i=1.2\times 10^7$, we choose $\Omega_{\phi i}=10^{-10}-10^{-6}$ and different $w_{\phi i}$.
For $\lambda_i=5.2\times 10^7$, we choose $\Omega_{\phi i}=10^{-2}-10^{-7}$ and different $w_{\phi i}$.
The $\times$ corresponds to the points $(w_{\phi 0},\Omega_{\phi 0})$
with $\lambda_i=1.2\times 10^7$ and different $\Omega_{\phi i}$ and $w_{\phi i}$.
The $+$ corresponds to the points $(w_{\phi 0},\Omega_{\phi 0})$
with $\lambda_i=5.2\times 10^7$ and different $\Omega_{\phi i}$ and $w_{\phi i}$. The dashed line is the fitting function (\ref{womgfit1}).
The purple and green dotted lines are for $\lambda_i=10$ without the tracking behavior.}
\label{pwrwomg}
\end{figure}

Because the dynamical system (\ref{dynaeq1a})-(\ref{dynaeq3a}) is hard to solve numerically if
$\lambda_i$ is too large, we choose to solve the dynamical system
(\ref{dynaeq1})-(\ref{dynaeq3}) numerically for the inverse power-law potential $V(\phi)=V_0(\phi/m_{pl})^{-6}$ to
illustrate the tracking behavior and the results are shown in Fig. \ref{pwrtrak1}. As seen
from Fig. \ref{pwrtrak1}, the general dynamics of the tracker field follows our discussion in the previous section,
and $w_\phi$ exhibits oscillatory behaviors before the scalar field reaches the tracking solution.
For small $\lambda_i=10$ (for this case, $\lambda_i\lesssim 100$), the tracking solution with constant $w_\phi$ does not appear
because $\Omega_{\phi}$ reaches $1$ in a short time.
Even though the general $w_\phi-\Omega_\phi$ relation was not followed
as shown by the dotted lines in Fig. \ref{pwrwomg}, the same $w_\phi-\lambda$
and $w_\phi-w_\phi'$ relations are still followed for those solutions with small $\lambda_i$
when $w_\phi$ approaches $-1$.
For large $\lambda_i$, the tracking behavior is realized easily.
But to get the observationally allowed $\Omega_{\phi 0}$ (the subscript $0$ means the current value),
we need to adjust the initial values of $\lambda$, $\Omega_\phi$ and $w_\phi$.
For the example shown in Fig. \ref{pwrtrak1}, we choose $\lambda_i=1.2\times 10^6$ and $10^{-17}\le \Omega_{\phi i}\le 10^{-11}$
so that $0.05\lesssim\Omega_{\phi 0}\lesssim 0.95$.
We also show the relation between $w_{\phi 0}$ and $\Omega_{\phi 0}$
for the tracking solutions with different initial conditions in Fig. \ref{pwrwomg}.
Not only the relation between $w_{\phi 0}$ and $\Omega_{\phi 0}$ exists, but
also the same relation follows for $w_\phi$ and $\Omega_\phi$ at any moment after it reaches the tracking solution.
This is one of the main results obtained in this Letter.

\begin{figure}[htp]
\begin{tabular}{cc}
\includegraphics[width=0.4\textwidth]{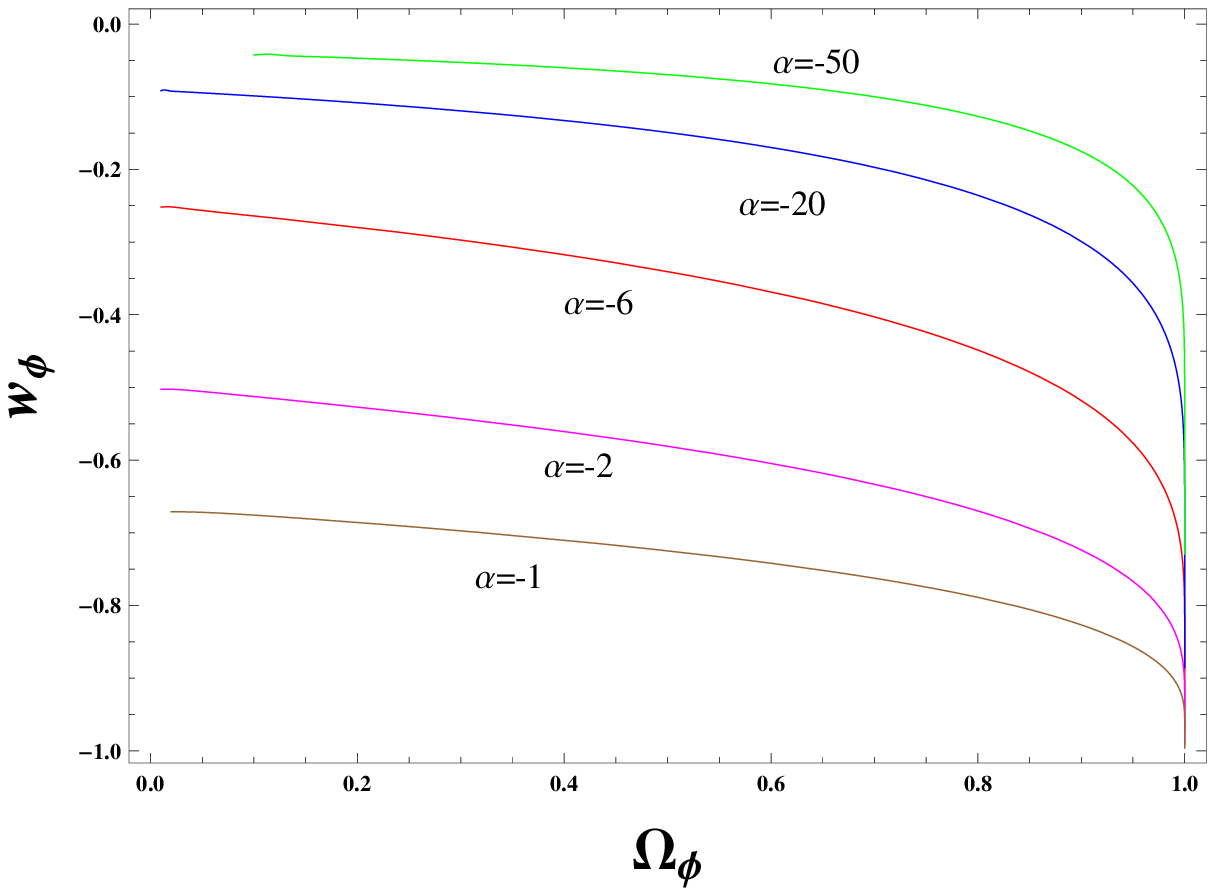}&\includegraphics[width=0.4\textwidth]{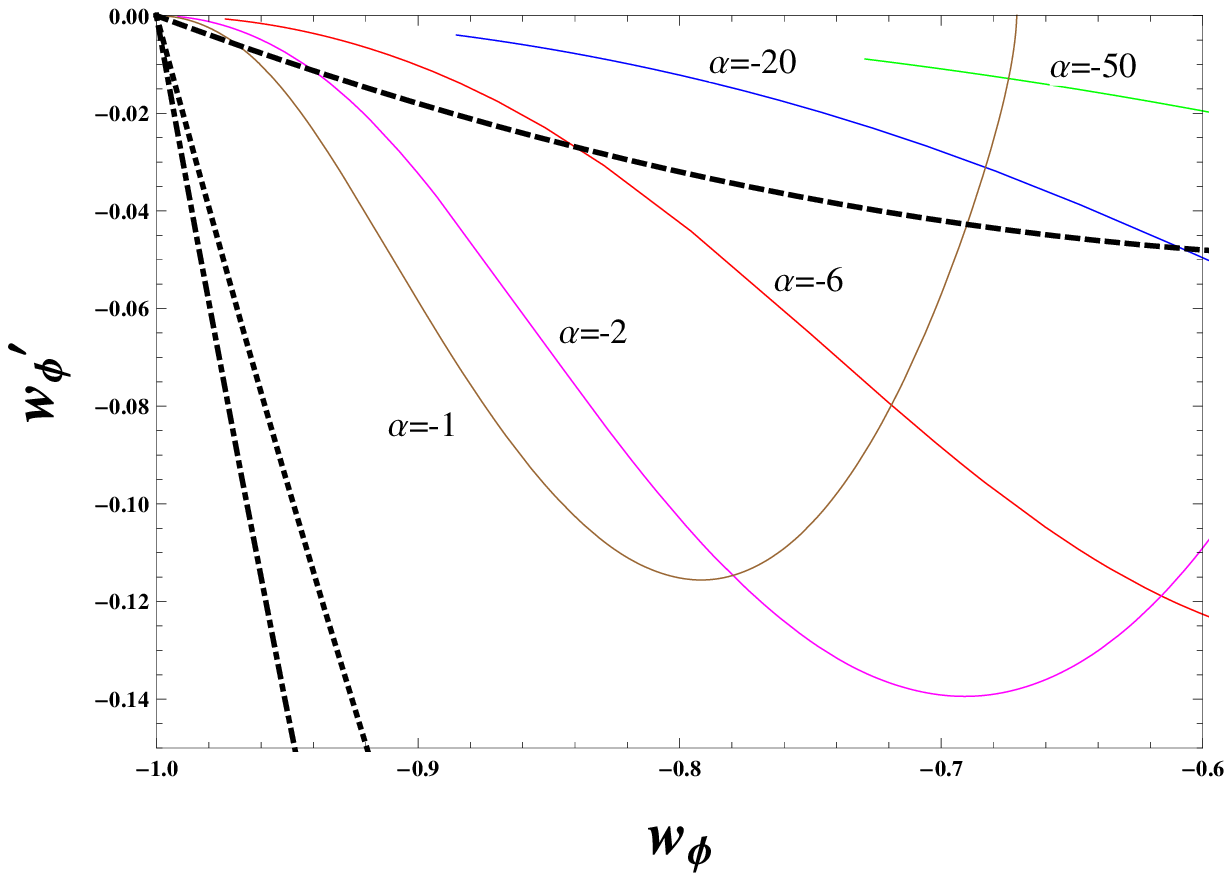}
\end{tabular}
\caption{The $w_\phi-\Omega_\phi$ and $w'_\phi-w_\phi$ relations
for the power-law potential $V(\phi)\sim \phi^\alpha$ with the tracking behavior.
The dashed line in the right panel is for the upper limit $0.2w_\phi(1+w_\phi)$, the dotted line is
for the tracking lower limit $3w_\phi(1-w_\phi^2)/(1-2w_\phi)$, and the dot-dashed line is for
the freezing lower limit $3w_\phi(1+w_\phi)$.}
\label{wwderfig}
\end{figure}

As we discussed in the previous section, the $w_\phi$-$\Omega_\phi$ trajectory is independent of the initial conditions,
so the $w_{\phi 0}$-$\Omega_{\phi 0}$ relation is the same as the general
$w_\phi$-$\Omega_\phi$ relation for the tracking solution because of the tracker condition,
although the values of $w_{\phi 0}$ and $\Omega_{\phi 0}$ depend on the initial conditions.
Therefore, we generalize the common $w_{\phi 0}-\Omega_{\phi 0}$ trajectory found in \cite{track1,track2}
to the common $w_{\phi}-\Omega_{\phi}$ trajectory for the tracking solutions even though
$w_\phi$ evolves differently when the tracker field starts to dominate the Universe.
The trajectory can be obtained by solving the dynamical system (\ref{dynaeq1a})-(\ref{dynaeq3a})
with the initial conditions (\ref{trkcond}) and (\ref{trksol2}).
A general $w_\phi$-$\Omega_\phi$ relation was proposed in \cite{Sourish} for slow-roll freezing
quintessence by assuming constant $\lambda$ as
\begin{equation}
\label{womgrel1}
\gamma_\phi=\frac{\lambda_0^2}{3}\left[\frac{1}{\sqrt{\Omega_\phi}}-\left(\frac{1}{\Omega_\phi}-1\right)(\tanh^{-1}(\sqrt{\Omega_\phi})+C)\right]^2.
\end{equation}
Apparently, this relation cannot be applied for the tracking behavior
because $\gamma_\phi\rightarrow \lambda_0^2/3$ when $\Omega_\phi\rightarrow 1$
and $\gamma_\phi\rightarrow \infty$ when $\Omega_\phi\rightarrow 0$ if $C\neq 0$.
The reason why the above relation does not work
is that $\lambda$ is not a constant for the tracking solution as shown in Fig. \ref{pwrwomg}.
When $\Omega_\phi$ is small, a linear approximation for the $w_\phi-\Omega_\phi$ relation was found in \cite{Watson:2003kk,Chiba:2009gg}.
We find that the general $w_\phi-\Omega_\phi$ relation for $\alpha=-6$ can be fitted by the following function
\begin{equation}
\label{womgfit1}
w_\phi=\frac{w_b(1-\Omega_\phi)+0.09\Omega_\phi+0.03\Omega_\phi^2-2(\Gamma-1)}{2\Gamma-1-\Omega_\phi}.
\end{equation}
It is obvious that $w_\phi$ does not differ much from the initial value (\ref{trksol2}),
so $w'_\phi$ is small. To see how small it is, we show the $w'_\phi-w_\phi$ trajectory
for different $\alpha$ in Fig. \ref{wwderfig}.
The upper limit $w'_\phi\lesssim 0.2w_\phi(1+w_\phi)$ \cite{Caldwell:2005tm} is also
shown in Fig. \ref{wwderfig}. Our results
show that the upper limit is violated. As we discussed above, as $w_\phi\rightarrow -1$, $F=1/3$
and $w_\phi'=0$,
so we expect the violation of the upper limit $0.2w_\phi(1+w_\phi)$.
The other problem is the observational constraints on the
values of $\Omega_{m0}$ and $w_{\phi 0}$.  Since $w_\phi$ and $\Omega_\phi$ follow a universal relation
which is independent of the initial conditions and the energy scale $V_0$ of the potential,
we can use the observational data to constrain the form of tracker potential.
For the power-law potential $V(\phi)=V_0(\phi/m_{pl})^\alpha$,
the observational constraints can be satisfied by choosing small $\alpha$ as shown in Fig. \ref{wwderfig}.
If we choose the observational constraints $0.2\le \Omega_{m 0}\le 0.4$ and $w_0\le -0.8$ \cite{gongmnras13,Gao:2013pfa},
then we require $0>\alpha\gtrsim -1$.

\subsection{Thawing solution}

\begin{figure}[htp]
\begin{tabular}{cc}
\includegraphics[width=0.4\textwidth]{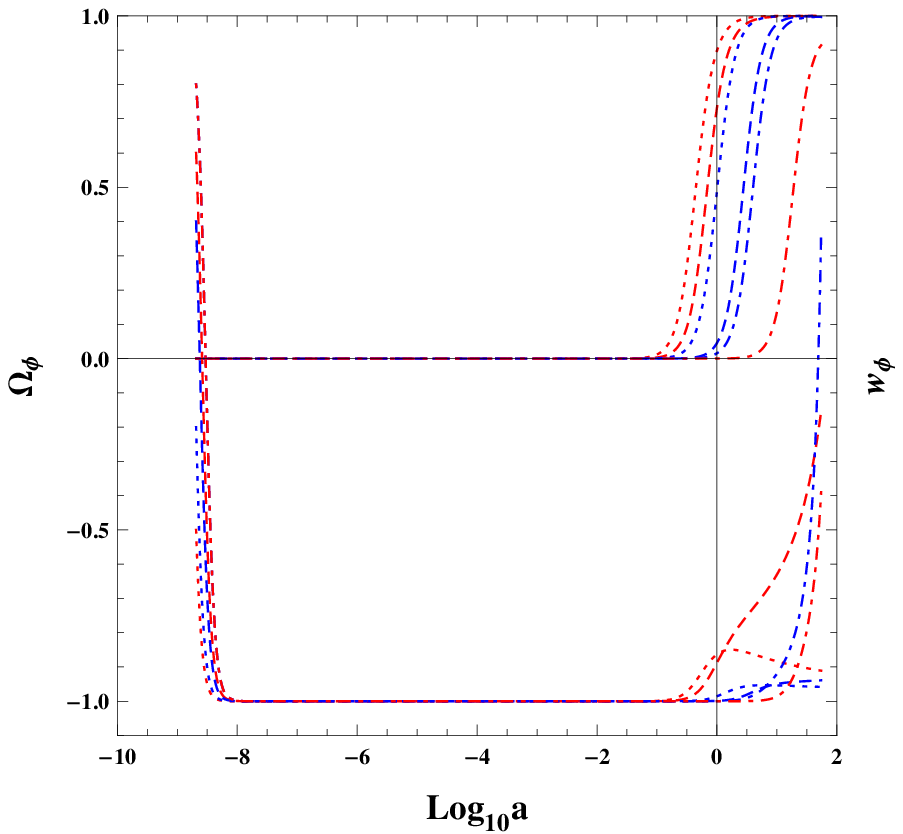}&\includegraphics[width=0.4\textwidth]{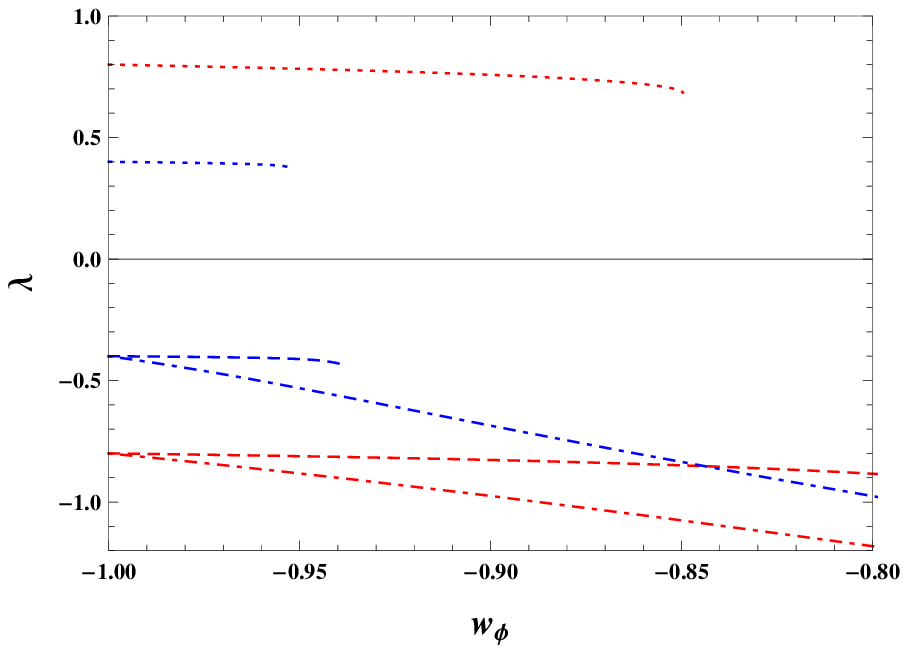}
\end{tabular}
\caption{The left panel shows the evolutions of $\Omega_\phi$ and $w_\phi$ and the
right panel shows the evolutions of $\lambda$ for the thawing solutions. The arbitrary initial time
$\ln a_i=-20$ was chosen for computational convenience. The red lines are for $|\lambda_i|=0.8$ and the blue lines are for $|\lambda_i|=0.4$.
The dashed lines are for the power-law potential $V(\phi)\sim \phi^6$, the dotted lines are for the inverse power-law potential $V(\phi)\sim \phi^{-4}$,
and the dot-dashed lines are for the PNGB potential.}
\label{thawevolfig}
\end{figure}

\begin{figure}[htp]
\begin{tabular}{cc}
\includegraphics[width=0.4\textwidth]{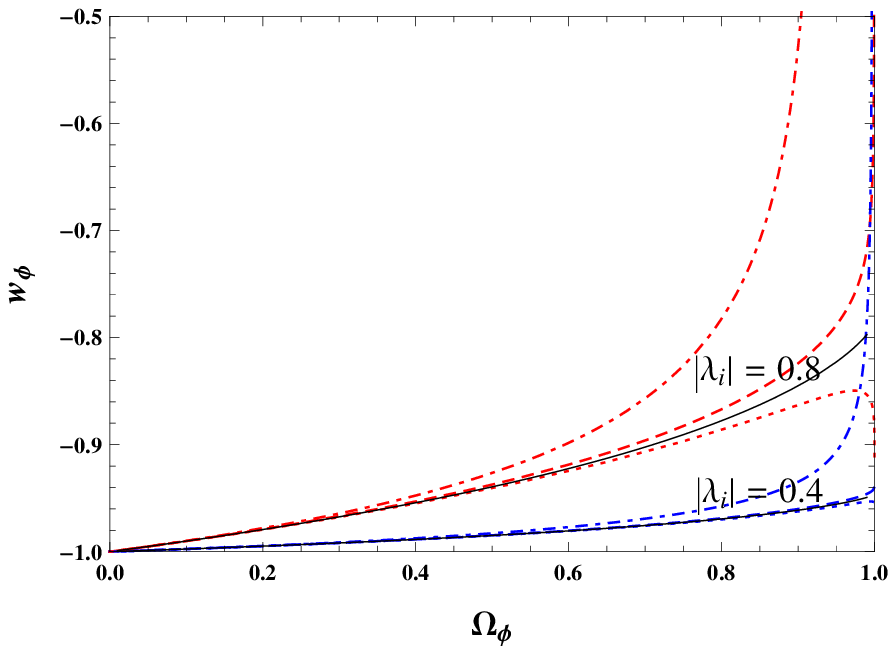}&\includegraphics[width=0.4\textwidth]{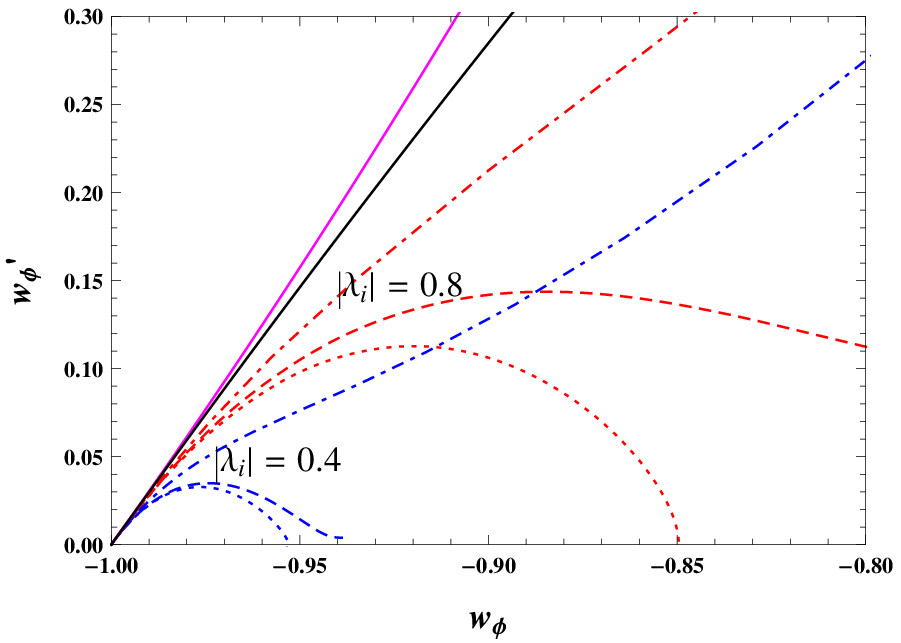}
\end{tabular}
\caption{The $w_\phi-\Omega_\phi$ and $w'_\phi-w_\phi$ relations for the thawing solutions.
The dashed lines are for the power-law potential $V(\phi)=\phi^6$, the dotted lines are for the inverse power-law potential $V(\phi)=\phi^{-4}$,
and the dot-dashed lines are for the PNGB potential.
The black lines in the left panel denote the analytical result (\ref{womgrel1}) with $C=0$.
In the right panel, the magenta line denotes the upper limit $3(1+w_\phi)(2+w_\phi)$ for thawing models,
and the black line denotes the upper limit $3(1-w_\phi^2)/2$.}
\label{thawrelfig}
\end{figure}

For the inverse power-law potential, if $\lambda_i$ is small, then $w_\phi$ decreases to $-1$ as seen from Eq. (\ref{dynaeq2a}).
If we also fine-tune the initial
value of $\Omega_{\phi}$ (for $\alpha=-4$, $\Omega_{\phi i}$ is around $10^{-30}$ at $\ln a_i=-20$), then $w_\phi$ stays
at the value $-1$ and starts to increase recently, we get the thawing behavior.
From Eq. (\ref{dynaeq3a}), it is easy to see that $\lambda$ will keep to be a constant when $w_\phi=-1$.
Combining Eqs. (\ref{dynaeq1a}) and (\ref{dynaeq2a}), we get
\begin{equation}
\label{womgeq2}
\frac{d\gamma_\phi}{d\Omega_\phi}=\frac{-3\gamma_\phi(2-\gamma_\phi)+\lambda(2-\gamma_\phi)\sqrt{3\gamma_\phi\Omega_\phi}}{3(\gamma_b-\gamma_\phi)\Omega_\phi(1-\Omega_\phi)}.
\end{equation}
Taking the approximation $\gamma_\phi\ll 1$, then Eq. (\ref{womgeq2}) can be approximated as
\begin{equation}
\label{womgeq3}
\frac{d\gamma_\phi}{d\Omega_\phi}=\frac{-6\gamma_\phi+2\lambda\sqrt{3\gamma_\phi\Omega_\phi}}{3\gamma_b\Omega_\phi(1-\Omega_\phi)}.
\end{equation}
The solution to the above Eq. (\ref{womgeq3}) with constant $\lambda\approx \lambda_i$ is
\begin{equation}
\label{womgrel3}
\gamma_\phi=\frac{\lambda_i^2}{3}\left(1+\frac{1}{2}\gamma_b\right)^{-2}\Omega_\phi(1-\Omega_\phi)^{2/\gamma_b}\,_2F_1^2
\left(\frac{1}{\gamma_b}+\frac{1}{2},\frac{1}{\gamma_b}+1,\frac{1}{\gamma_b}+\frac{3}{2};\Omega_\phi\right),
\end{equation}
where $_2F_1(a,b,c,x)$ is the hypergeometric function. This approximation breaks down when $\gamma_\phi \sim 1$.
As $\Omega_\phi\rightarrow 0$ and $w_\phi\rightarrow -1$,
$\gamma_\phi\rightarrow \lambda_0^2\Omega_\phi/3(1+\gamma_b/2)^2$, so the flow parameter $F=1/3(1+\gamma_b/2)^2$
and $\beta=\gamma_b/2$ which is consistent with the result found in \cite{Chiba:2009sj} with different argument.
If $w_\phi$ starts to increase during the matter domination,
$\gamma_b=1$ and $F=4/27$, we recover the familiar $w_\phi-\Omega_\phi$
relation (\ref{womgrel1}) with $C=0$. We
show the evolutions of $\Omega_\phi$, $w_\phi$ and $\lambda$ in Fig. \ref{thawevolfig},
and the $w_\phi-\Omega_\phi$ and $w_\phi-w'_\phi$ relations
are shown in Fig. \ref{thawrelfig} with dotted lines for the inverse power-law
potential with $\alpha=-4$.
We choose two different initial values of  $\lambda_i=0.8$ and $\lambda_i=0.4$. The thawing solution was kept
up to $w_\phi\sim -0.95$ for $\lambda_i=0.4$ and $w_\phi\sim -0.85$ for $\lambda_i=0.8$. When the scalar field takes
the thawing solution, $\lambda$ is almost a constant as shown in Fig. \ref{thawevolfig}
and the analytical relation (\ref{womgrel3}) approximates
the $w_\phi-\Omega_\phi$ relation well as shown in Fig. \ref{thawrelfig}.
Since $\lambda'\propto \lambda^2$, the larger $\lambda$ is,
the faster $\lambda$ changes, so the analytical relation (\ref{womgrel3}) gives better approximation for smaller $\lambda_i$
as shown in Fig. \ref{thawrelfig}.

For the power-law potential with positive $\alpha$, the roll parameter $|\lambda|$ increases with time and
there is no asymptotically freezing solution. To get the thawing solution,
we need to start with small $|\lambda_i|$
so that $w_\phi$ decreases to $-1$.
If we also fine-tune the initial
value of $\Omega_{\phi}$ (for $\alpha=6$, $\Omega_{\phi i}$ is around $10^{-30}$ at $\ln a_i=-20$), then $w_\phi$ stays
at the value $-1$ and starts to increase recently.
We show the evolutions of $\Omega_\phi$, $w_\phi$ and $\lambda$ in Fig. \ref{thawevolfig},
the $w_\phi-\Omega_\phi$ and $w_\phi-w'_\phi$ relations in Fig. \ref{thawrelfig} by
the dashed lines for the power-law potential with $\alpha=6$.
We choose two different initial values of  $\lambda_i=-0.8$ and $\lambda_i=-0.4$.
When the scalar field takes
the thawing solution, $\lambda$ is almost a constant as shown in Fig. \ref{thawevolfig}
and the analytical relation (\ref{womgrel3}) approximates
the $w_\phi-\Omega_\phi$ relation well as shown in Fig. \ref{thawrelfig}.
Again the analytical relation (\ref{womgrel3}) gives better approximation for smaller $|\lambda_i|$
as shown in Fig. \ref{thawrelfig}.
If we use the observational constraints $w_{\phi 0}\le -0.8$ and $\Omega_{\phi 0}>0.6$,
then the analytical relation (\ref{womgrel1}) requires $|\lambda_i|<1.3$.
From the analytical relation (\ref{womgrel1}), we see that $\lambda\rightarrow 3(1+w_\phi)/\Omega_\phi$
as $\Omega_\phi\rightarrow 1$ which is not true for the positive power-law potential
because $|\lambda|$ keeps increasing and it increases faster and faster once $w_\phi$ deviate from $-1$, this means that
the approximation is broken as $\Omega_\phi\rightarrow 1$. On the other hand, as $\Omega_\phi\rightarrow 0$,
we get $1+w_\phi\rightarrow \lambda^2\Omega_\phi/3(1+\gamma_b/2)^2=4\lambda^2\Omega_\phi/27$,
so the flow parameter $F=1/3(1+\gamma_b/2)^2=4/27$ initially at the matter domination
for the thawing solution and the flow parameter $F=1/3$ when the quintessence field leaves the thawing solution.
Therefore, $4/27\le F\le 1/3$ for the thawing solution, we get an upper limit $w_\phi'\le 3(1-w_\phi^2)/2$
which is smaller than the upper limit $w_\phi'=3(1+w_\phi)(2+w_\phi)$ \cite{Caldwell:2005tm},
and there is no lower limit on $w_\phi'$. As shown in Fig. \ref{thawrelfig},
the lower limit $w_\phi'\ge 1+w_\phi$ \cite{Caldwell:2005tm} does not hold.
These two upper bounds are also shown in Fig. \ref{thawrelfig} and they are satisfied by the thawing solutions.
From Eq. (\ref{betaeq1}), we get $\beta=\gamma_b/2=1/2$ initially and $\beta=-\gamma_\phi/2$ when
the thawing period ends.

\section{PNGB potential}

In this section, we focus on the $w_\phi-\Omega_\phi$ approximation (\ref{womgrel3})
and the limit on $w_\phi'$ for the PNGB potential.
The PNGB potential $V(\phi)=M^4[1\pm \cos(N\phi/f)]$ was first
proposed in the schizon model in which the small PNGB mass
is protected by fermionic chiral symmetries \cite{Hill:1987bm,Hill:1988bu},
$\pi$ meson and the axion are examples of PNGB. In cosmology, the PNGB potential was first
introduced as natural inflation \cite{Freese:1990rb}, and was later found that it also dominates the
energy density of the universe at present \cite{Frieman:1995pm}.
In this Letter, we choose the PNGB potential without loss of generality, $V(\phi)=M^4[1-\cos(\phi/m_{pl})]$
with the energy scale $M\sim 10^{-3}$eV,
the tracker parameter $\Gamma=(\lambda^2-1)/2\lambda^2<1$
and $f(\lambda)=\Gamma-1=-(1+\lambda^2)/2\lambda^2$ \cite{Fang:2008fw},
so there is no tracking solution for the PNGB potential. Since $f(0)$ is not well defined,
the dynamical analysis on the fixed points in \cite{Fang:2008fw} is not applicable.
The critical points are: $\Omega_{\phi c}=0$, $\gamma_c=0$
or $\gamma_c=2$ with arbitrary $\lambda$; $\Omega_{\phi c}=1$ and $\gamma_c=0$ with arbitrary $\lambda$; and
$(\Omega_{\phi c},\gamma_c,\lambda_c)=(1,2,0)$.
In fact, all the critical points are unstable points.

In this model, $\lambda'\propto (1+\lambda)^2$, the roll parameter $\lambda$ changes faster than the power-law potential,
so we don't expect that the analytical expression (\ref{womgrel3}) approximates the $w_\phi-\Omega_\phi$
as well as that for the power-law potential with the same $\lambda_i$.  To get the thawing solution,
we also need to start with small $|\lambda_i|$ so that $w_\phi$ quickly reaches the initial thawing value $-1$,
and we also need to fine-tune the initial
value of $\Omega_{\phi}$ to be around $10^{-32}$ at $\ln a_i=-20$.
The evolutions of $\Omega_\phi$, $w_\phi$ and $\lambda$ are shown in Fig. \ref{thawevolfig},
and the $w_\phi-\Omega_\phi$ and $w_\phi-w'_\phi$ relations are shown in Fig. \ref{thawrelfig}
with dot-dashed lines. As we expect, $\lambda$ does not keep to be nearly constant,
the approximation (\ref{womgrel1}) is not good for large $\lambda_i$ and
it breaks down when $\Omega_\phi$ approaches 1.

\section{Discussion and Conclusions}

When the tracking solution is reached, $w_\phi'\approx w_\phi^{''}\approx 0$, so $w_\phi$ is almost a constant
and both the tracker condition (\ref{trkcond})
and Eq. (\ref{trksol2}) are satisfied. To keep $w_\phi$ to be a constant, $\Omega_\phi$ should be small
and the tracker parameter $\Gamma$ should  be nearly constant, so the
tracker condition (\ref{trkcond}) requires the roll parameter $\lambda$ to be large. Therefore,
the tracker parameter $\Gamma>1$ and large initial value of the roll parameter $\lambda$
are the the necessary conditions for tracking solutions.
Based on this analysis, we proposed the tracker theorem.
Although the current value of $\Omega_\phi$ and $w_\phi$ depend on
the initial conditions for the tracking solutions, the $w_\phi-\Omega_\phi$ trajectory
before $\Omega_\phi$ reaches 1 is
independent of the initial conditions and it can be used to exclude models by comparing it
with the observational constraints. If we choose the observational
constraints $0.2\le \Omega_{m 0}\le 0.4$ and $w_0\le -0.8$,
then we require $n\lesssim 1$ for the inverse power-law potential $V(\phi)=V_0(\phi/m_{pl})^{-n}$.
Since the dark energy domination ($\Omega_{\phi}=1,\ w_\phi=-1,\ \lambda=0$)
is the attractor for the inverse power-law potential, the asymptotic behaviors of $\lambda$ and $w_\phi$
are the same and the same asymptotic $\lambda-w_\phi$ trajectory is followed by all solutions including the
tracking and non-tracking solutions. The flow parameter $F$ starts and ends with $F=1/3$, the upper bound
$0.2w_\phi(1+w_\phi)$ does not hold and we expect that no such upper bound exists for the freezing models,
this will make the distinction between cosmological constant and dynamical tracker fields more difficult.

If the initial value of the roll parameter $\lambda$ is small, then $w_\phi$ quickly decreases to -1 and stays at the value
until the roll parameter $\lambda$ becomes large, after that $w_\phi$ starts to increase. This thawing behavior can be achieved for
the power-law potential with positive $\alpha$ and the the PNGB potential. The thawing behavior can also be achieved
for the inverse power-law potential for a period of time if the initial value
of the roll parameter is small. In general, we need to fine-tune the initial conditions so
that we get the right values of $\Omega_{\phi 0}$ and $w_{\phi 0}$ which are consistent with the observational constraints
for the thawing solutions. Because $w_\phi\approx -1$ initially, the roll parameter changes very slowly and
it can be approximated as a constant,
a general $w_\phi-\Omega_\phi$ relation (\ref{womgrel3}) is then obtained. Based
on the asymptotical behavior of the $w_\phi-\Omega_\phi$ relation, the flow parameter $F=1/3(1+\gamma_b/2)^2=4/27$
when $\Omega_\phi\rightarrow 0$ and $w_\phi\rightarrow -1$ during the matter domination,
and $F=1/3$ when the thawing behavior ends,
we derive the upper bound $w_\phi'\le 3(1-w_\phi^2)/2$
and we expect that no lower bound exists for the thawing models,
so the distinction between cosmological constant and dynamical thawing models becomes more difficult.
If we use the observational constraint $w_{\phi 0}<-0.8$ and $0.2<\Omega_{m0}<0.4$,
we find that the initial value of the roll parameter $|\lambda_i|<1.3$ for the potentials with the thawing solutions.

In summary, we find that the same relation not only exists between $w_{\phi 0}$ and $\Omega_{\phi 0}$,
but also exists between $w_\phi$ and $\Omega_\phi$ at any time after the tracker field takes the tracking solutions.
The relation is independent of the initial conditions and the energy scale $V_0$ of the tracker field, so the
observational data can be used to constrain the tracker model by using this relation.
Based on the existence of the relation, we generalize the tracking solutions with a common track of $w_\phi(t)$
to those solutions with a common $w_\phi-\Omega_\phi$ trajectory and we propose the tracker theorem by using
the roll parameter $\lambda$. Both the upper limit $w_\phi'< 0.2w_\phi(1+w_\phi)$ for the tracking solutions
and the lower limit $w_\phi'>1+w_\phi$ for the thawing solutions are found to
be violated, and we propose a lower upper bound $w_\phi'\le 3(1-w_\phi^2)/2$ for the thawing solutions.

\begin{acknowledgments}
This work was partially supported by
the National Basic Science Program (Project 973) of China under
grant No. 2010CB833004, the NNSF of China under grant Nos. 10935013 and 11175270,
the Program for New Century Excellent Talents in University under grant No. NCET-12-0205
and the Fundamental Research Funds for the Central Universities under grant No. 2013YQ055.
\end{acknowledgments}

\end{document}